# The new insight into gallium nitride (GaN) melting under pressure

[1]Sylwester Porowski, [1,3]Bogdan Sadovyi, [1]Stanisław Gierlotka, [1,4]Sylwester J. Rzoska,

[1]Izabella Grzegory, [2]Igor Petrusha, [2]Vladimir Turkevich, [2]Denys Stratiichuk

[1]Institute of High Pressure Physics "Unipress", Polish Academy of Sciences, Sokołowska str. 29/37, 01-142 Warsaw, Poland

[2]V. Bakul Institute for Superhard Materials of the National Academy of Sciences of Ukraine, Avtozavodska str. 2, Kyiv 04074, Ukraine

[3]Department of Physics, Ivan Franko National University of Lviv, Dragomanova str. 50, Lviv UA 79005, Ukraine

[4]ŚMCEBI & Institute of Physics, University of Silesia,. ul. 75 Pułku Piechoty 1, 41-500 Chorzów, Poland

**Abstract**

Results solving the long standing puzzle regarding the phase diagram and the pressure evolution of the melting temperature $T_m(P)$ of gallium nitride (GaN), the most promising semiconducting material for innovative modern electronic applications, are presented. The analysis is based on (**i**) studies of the decomposition curve in *P-T* plane up to challenging $P \approx 9 GPa$, (**ii**) novel method enabling $T_m(P)$ determination despite the earlier decomposition, and (**iii**) the pressure invariant parameterization of $T_m(P)$ curve, showing the reversal melting for $P > 22$ GPa. This is linked to a possible fluid – fluid crossover under extreme pressures and temperatures. The importance of results for the development of GaN based technologies is indicated.

# 1. Introduction

Silicon, germanium as well as $A^{III}B^V$ and $A^{II}B^{VI}$ semiconductors belong to the group of the most important functional materials for the nowadays civilization.[1, 2] In the 21$^{st}$ century the exceptional success reached gallium nitride (GaN), which applications revolutionized innovative light sources[3, 4] and high power and high frequency electronics[5, 6]. The market of GaN-based devices has already reached the second position after dominating silicon one, but still there is a significant barrier in its development due to the very limited access to high quality bulk single crystals. They cannot be grown using classical Czochralski or Bridgeman methods from stoichiometric melts,[12] since GaN decomposes prior to melting. So far, the highest quality GaN crystals were grown under high pressure, $P \sim 1$ GPa, from the nitrogen solution in gallium. However, due to still low solubilities (< 1 at. %) the rate of crystallization remained low.[13] Notably higher values (>10 at. %) reported below indicate new possibilities for obtaining perfect and large GaN crystals.

The elemental group IV and binary semiconductor compounds are tetrahedrally coordinated (zinc blende or wurtzite) low density covalent crystals. It was believed that all of them, including GaN, melt into higher density metallic liquids with the coordination number $CN \approx 6$, what results in decreasing of melting temperature ($T_m$) with rising pressure ($P$), i.e. $dT_m/dP < 0$. This general picture was described by Van Vechten[7] in his *Quantum Dielectric Theory of Chemical Bonding*, still constituting one of basic references for the physics of semiconductors. Nevertheless, in the last decade experimental[8] and theoretical[9] evidences disturbing this general picture appeared. For liquid germanium with local CN≈6 notable fluctuations with the solid like coordination CN≈4 were detected.[8] For liquid GaN, the local CN≈4, the same as in the solid phase, was predicted by Harafuji et al.[9] via Molecular Dynamics simulations. This lead to the increase of $T_m$ with compressing, i.e. $dT_m/dP > 0$.[9] The situation became even more puzzling following the esteemed experimental report by Utsumi et al.[10], who indicated that GaN decomposes up to $P \approx 5.8 GPa$ and subsequently it melts at constant temperature $T_m \approx 2500K$ up to at least $P \approx 10 GPa$, what is related to $dT_m/dP \approx 0$.

To solve this Gordian knot of contradictory results, the experimental evidence showing that GaN decomposes prior to melting up to at least $P$ = 9 GPa and $T$ = 3400 K is presented. High pressure – high temperature (*HP-HT*) experiments was carried out in extraordinary large and does not employed so far pressurized volume $V \sim 1cm^3$, essentially minimizing parasitic, disturbing artefacts. Results were analysed via the innovative scaling protocol determining loci of $T_m(P)$ despite decomposition occurring in prior. The novel, pressure invariant, way of $T_m(P)$ parameterization indicated its possible maximum at ca. 22 GPa, presumably associated with the structural liquid-liquid transition/crossover.

## 2.    Results and discussion

### 2.1 Determining of High pressure and High temperature stability of GaN

The focus of experimental studies discussed below was the minimization of the impact of parasitic artefacts, often disturbing *P-T* phase diagram studies carried out so far. The range of pressures and temperatures was extended above the limit tested so far. To reach these targets high pressure – high temperature, up to *P* = 9 GPa and *T* = 3400 K, experiments were performed in the pressure set-up with a relatively large pressurized volume allowing for ca. 1 cm³ samples (Fig. 1)[14]. This made it possible to use large and high structural quality GaN single crystals[15] for melting and decomposition experiments. All these reduced the parasitic impact of the interactions between the bulk sample and the high pressure container, leading to the superior characterization of *HP-HT* induced processes in GaN samples. Twenty *HP-HT* experiments were performed. Properties of HP-HT treated samples, were analysed by High Resolution X-ray Diffraction (HR XRD) and powder X-ray Diffraction (XRD) supported by the scanning electron microscopy (SEM) and the optical stereo microscopy. The comparison of properties before and after the *HP-HT* treatment was conducted.

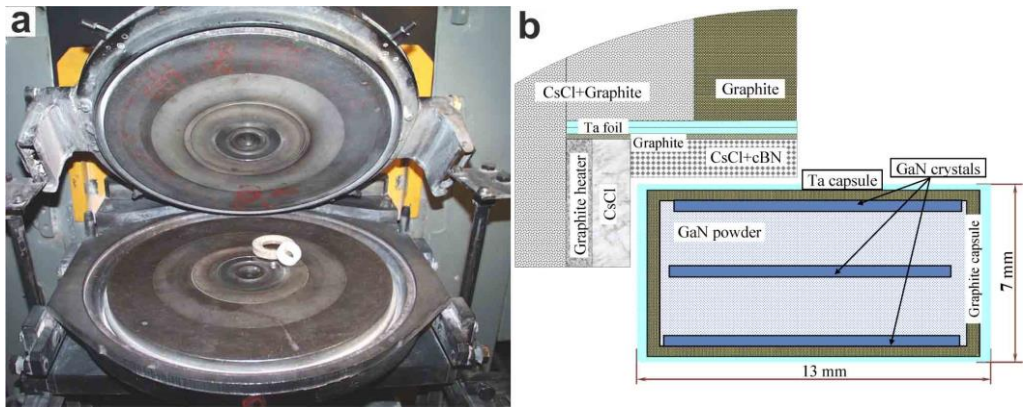

**Figure 1**    Toroid type apparatus for HP-HT studies: (a) photo of the pressurized part of the apparatus with the cell, and (b) the scheme of cross-section of *HP-HT* cell with graphite heater and the sample capsule.

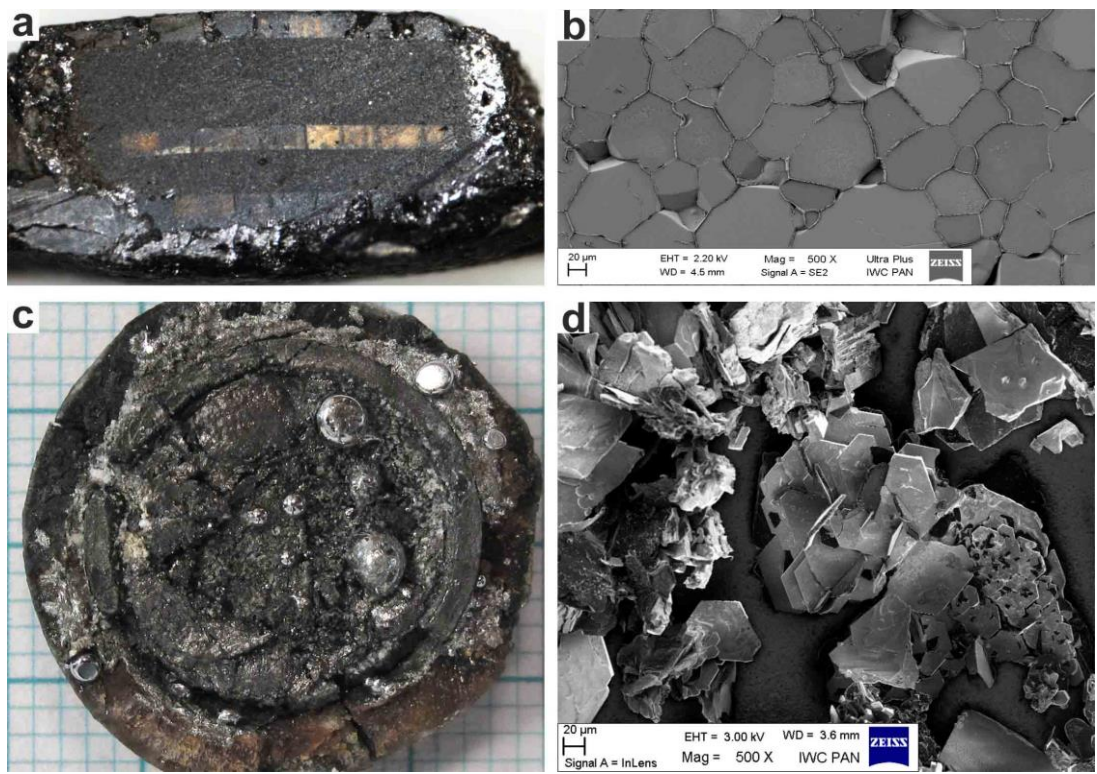

**Figure 2**    Optical microscopy and SEM views of GaN samples after *HP-HT* treatments revealing two types of behaviour. Type I: (a) there are no visible changes in the shape of GaN crystals and (b) there are hallmarks of GaN powder sintering. Type II: (c) the whole sample is decomposed and gallium droplets are visible, and (d) after removal (etching) of gallium newly grown crystals appear.

Typical results of the *HP-HT* annealing experiments are shown in Figure 2. Generally, two patterns of results have been obtained (Fig. 2). For type I results, there are no visible changes in the shape of GaN crystals after *HP-HT* treatment (Fig. 2a). However, several cracks due to some inhomogeneities of pressure, what is the inherent feature of quasi-hydrostatic solid medium pressure set-ups, are visible. The HR XRD analysis showed that GaN platelets remained single crystals with pressure induced misorientation of grains smaller than 30′ (see Supplementaries).

For the powder part of the sample sings of sintering were noted (Fig, 2b). However, the average size of powder grains did not increase substantially. All these showed that at HP-HT of type I experiments, gallium nitride was stable. It neither decomposed nor melted. For type II, GaN crystals and powder were totally decomposed, as seen in the optical microscopy image of the open capsule, where only liquid gallium is visible (Fig. 2c). Employing nitric acid based etching, gallium was removed and then GaN crystals in the form of thin hexagonal platelets, previously immersed in liquid gallium, appeared. These crystals are visible in the SEM image as shown in Figure 2d. The powder XRD analysis confirmed that solely GaN crystals have been obtained. Hence, for type II experiments GaN was unstable and it decomposed into $N_2$ gas and liquid Ga+$N_{atomic}$ solution. The observed newly grown crystals in the form of hexagonal platelets grew during cooling of supersaturated Ga+$N_{atomic}$ solution (Fig. 2d). The same type of GaN crystals was observed in previous experiments where Ga+$N_{atomic}$ solution was cooled for pressures below 1.5 GPa Grzegory et al.[16] Results obtained from type I (stable GaN) and type II (unstable GaN) experiments yielded the base for the extension the GaN equilibrium curve up 9 GPa, shown in Figure 3 below. They are is agreement with earlier results for pressures below 6 GPa by Karpinski et al.[17] for $P < 6 GPa$.

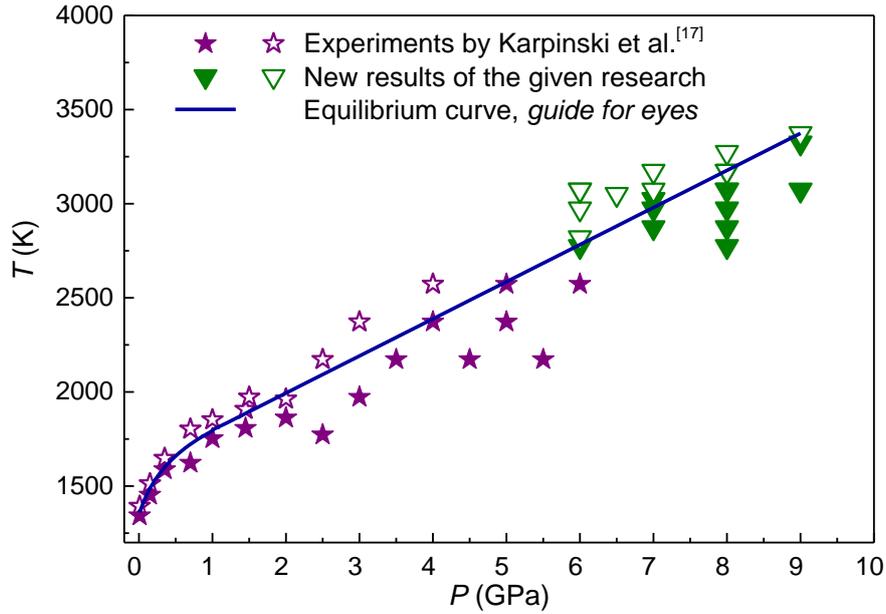

**Figure 3** GaN decomposition (EQ) curve. Experimental equilibrium curve for the solid GaN, liquid Ga+N$_{atomic}$ solution and gaseous N$_2$ in the pressure–temperature plane. ▲ ★ – GaN stable crystals, △ ☆ – GaN decomposed to (Ga+N)$_{liq.}$ and N$_2$, ─── the "guide for eyes" interpolation equilibrium curve.

## 2.2 Solubility of Nitrogen in Liquid Gallium

The crystallization of new GaN crystals in samples for which GaN decomposed during the *HP-HT* treatment allowed for the determination of the solubility of nitrogen in liquid gallium. The estimation of solubility under very high pressure and temperature is an extremely difficult experimental task. Grzegory et al.[16] in a set of brilliant experiments carried up to *P* = 1.5 GPa found that the reliable estimation of solubility in GaN is possible in the immediate vicinity of the stability curve in the *P-T* plane. In the present paper this finding were implemented. Following ref.[16] the system composed of liquid Ga + N solution and N$_2$ gas at temperatures and pressures corresponding to the *loci* just above the equilibrium *P-T* curve was equilibrated, so at this starting point GaN crystals were absent. Then the temperature was abruptly decreased to the room value and pressure was released. During cooling of the system the solution became supersaturated and small GaN hexagonal platelets crystallized. It was tested that gallium did not contain nitrogen after such process. The mass of the newly grown GaN crystals did not depend on the rate of cooling. Therefore, they contained all nitrogen previously dissolved in gallium. The knowledge of the mass of newly grown GaN crystals and the mass of gallium, made the determination of the solubility of nitrogen in gallium possible[16]. The solubility was calculated via:

$$S = \frac{n_N}{n_N + n_{Ga}} \times 100\% = x \times 100\% .\qquad(1)$$

where $n_{Ga}$ and $n_N$ denote the quantity (number of atoms) of gallium and nitrogen in the tested sample after decomposition and cooling.

For type II experiments values of $n_{Ga}$ and $n_N$ were calculated from measured masses of gallium and GaN newly grown crystals after the *HP-HT* treatment. Results are given in Table I. In col. 3 gives values of solubility from the given studies and from refs.[16, 18] In ref.[18] Nord et al. calculated the solubility of nitrogen in gallium for $P = 20$ GPa using MD and Monte Carlo simulations.

## 2.3 Scaling Analysis of Solubility for $T_m(P)$ Determination

GaN belongs to $A^{III}B^V$ homologous series of semiconductors, together with GaP, GaAs and GaSb. For GaP and GaSb the solubility curve $T(x)$ can be determined from relatively simple experimental tests under near-atmospheric pressure[19]. For these compounds solubilities of antimony and phosphorus in liquid gallium can be well portrayed by the ideal solution model[20], with approximately the same values of the melting entropy $\Delta S_m \approx$ 12.7 cal/(mole K)) (GaP) and 12.3 cal/(mole·K) (GaSb). They are shown in Figure 4a. GaAs does not follow this pattern because of specific properties of its liquid phase associated with the tendency of As aggregation[21].

For GaN under atmospheric pressure the nitrogen solubility is extremely low due to its decomposition of GaN at high temperatures[16]. At higher pressures the temperature range of GaN stability increases and consequently the maximum value of solubility rises. It was measured up to $P = 1.5$ GPa and $T = 1850$ K in ref.[16]. However, the ideal solution model based analysis of solubility at *HP-HT* conditions requires the knowledge of the pressure dependence of the melting temperature $T_m(P)$.

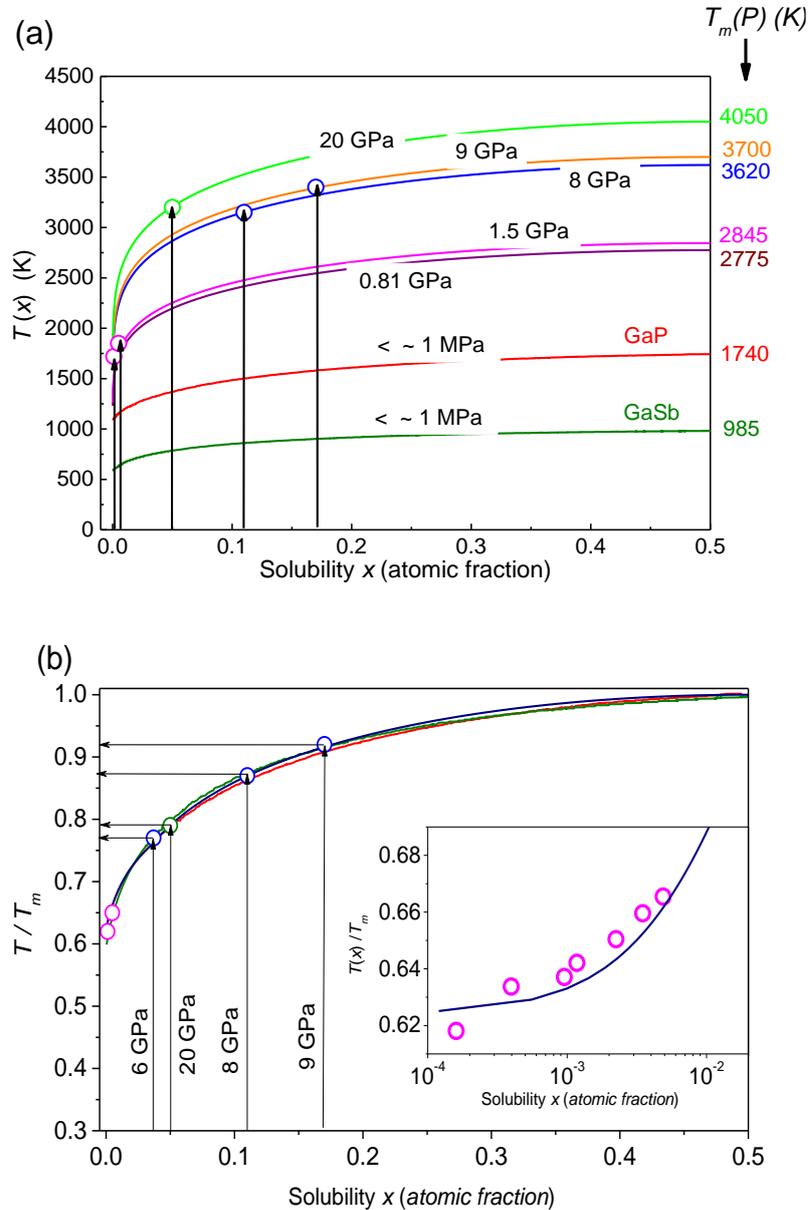

**Figure 4** Liquidus equilibrium (EQ) curves for GaSb, GaP and GaN for different pressures (a) and scaled liquidus curves for these compounds (b). ▬ is for GaSb, ▬ for GaP (based on data from ref.[21]) and ▬ for GaN from this work. ○ – Experimental solubility values obtained by Grzegory et al.[11], ○ – theoretical solubility value obtained by Nord et al.[18], ○ – experimental solubility values from this work.

As shown in Figure 4 the experimental solubility of nitrogen in gallium can be well portrayed by the ideal solution model with $\Delta S_m \approx 12.5$ cal/(mole K), confirming the similarity to solutions of phosphorus and antimony in gallium (GaP and GaSb cases). However, there is a basic difference in the behaviour of solubilities for GaP, GaSb and GaN cases. For GaN the solubility can be measured only at *HP-HT* conditions. For each pressure there is a different solubility curve (Fig. 4a) but in practice the reliable experimental determination of the solubility is possible only for one point each curve, related to the immediate vicinity of the equilibrium curve, as discussed above[16].

At first sight these experimental solubility data are chaotically scattered in the *T–x* plane (Fig. 4a). This scatter is inherently associated with the fact that experimental solubility data are for different pressures. However, the scatter disappears when introducing the scaled temperature $T/T_m(P)$, as shown in Figure 4b. Such way of analysis results also from the ideal solution model dependence[20]:

$$\frac{T}{T_m}(x) = \frac{\Delta S^m}{\Delta S^m - R \ln 4x(1-x)}. \qquad (2)$$

where $\Delta S_m$ is the entropy of melting, $x$ denotes the concentration of $B^V$ component and $R$ is for the gas constant.

This behaviour is shown in Figure 4b. All solubility can be approximately superposed into one scaled liquidus curve. Now, experimental solubility data are ordered and located on the scaled liquidus curve. It is also visible at very low solubilities, $x \leq 0.005$, where experimental data from[16] are well portrayed by the scaled liquidus curve (Fig. 4b, insert). The scaling protocol made it possible to determine parameters for columns 4 and 5 in Table I, yielding the experimental pressure dependence of the melting temperature for GaN.

## 2.4 Pressure Dependence of Melting Temperature $T_m(P)$

The most often used tool for portraying $T_m(P)$ evolution is the Simon-Glatzel (SG) equation[22,23] which predicts the permanent increase of $T_m$ with rising pressure, i.e. yielding always $dT_m/dP > 0$. However, the SG equation is inherently unable to portray the increasing group of materials with the maximum of $T_m(P)$ curve[24, 25] as well as materials where $dT_m/dP < 0$, such as germanium[26] or silicon[27].

The coherent description of all mentioned experimental types of melting curves is possible via the recently proposed semi-empirical equation by Drozd-Rzoska (DR):[11]

$$T_m(P) = T_{ref.}\left(1 + \frac{\Delta P}{\pi + P_{ref.}}\right)^{1/b} \times \exp\left(\frac{\Delta P}{c}\right), \text{ for } \Delta P = P - P_{ref.}, \qquad (3)$$

where $-\pi$ is the extrapolated, negative pressure for which $T_m(P \to -\pi) \to 0$.

For $(P_{ref.}, T_{ref.})$ one can assume arbitrary value along the melting curve. Pressure invariant values of all parameters in eq. (3) can be found from the plot based on the derivative equation resulting from eq. (3):[11]

$$\left[d(\ln T_m)/dP + c^{-1}\right]^{-1} = A + BP, \qquad (4)$$

where $A$ and $B$ can be determined from the linear regression analysis and then $\pi = A/B$ and $b = B$ (Fig. 5a).

The nonlinearity shown in the inset in Figure 5a indicates the inadequacy of the SG equation[22] for GaN. Basing on obtained values of coefficients $\pi$, $b$, $c$ given in Figure 5a one can portray $T_m(P)$ experimental data via eq. (3) (black stars and red curve in Fig. 5b) withput any additional fitting. Notable is the excellent agreement with results of MD modelling by Harafuji et al.[9] (open circles in Fig. 5b). Notable is the correlation with the melting temperature for $P = 20$ GPa estimated via the scaling analysis using the theoretical value of solubility calculated by Nord et. al.[18].

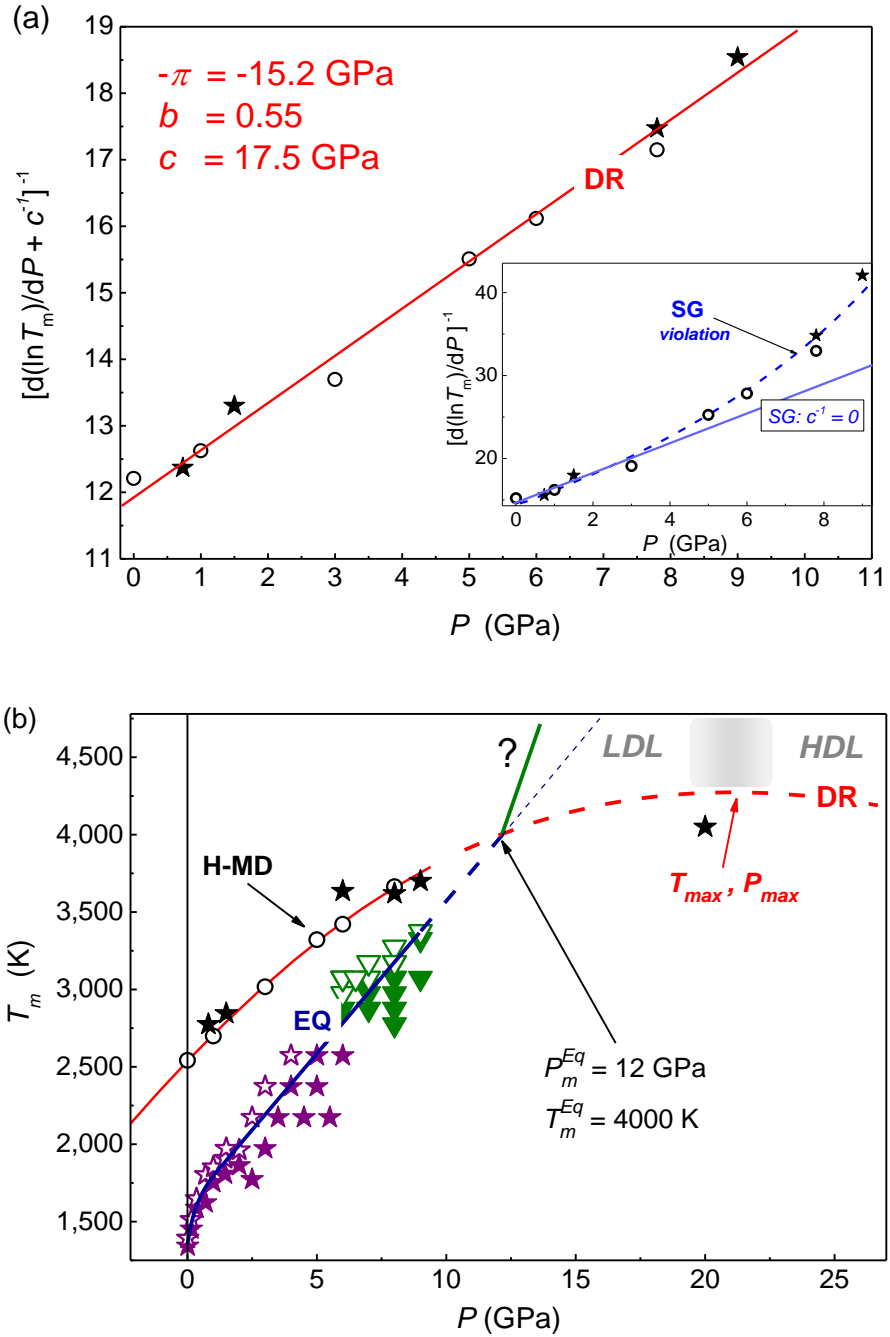

**Figure 5** Pressure dependence of GaN melting and GaN *P-T* equilibrium curve (EQ). a, shows results of the derivative based analysis, via eqs. (4) showing the failure of the SG[22] eq. (3), which validity ought to appear via the linear domain. Such behaviour is presented in the inset, where DR[11] eq. (3) indicated the possibility of parameterization via DR[11] eq. (4) is shown. The linear regression fit yields optimal values of all relevant coefficient ($\pi, b, c$) for eq. (3). $T_m(P)$ curve developed on this base is shown in the main part of b. ○ is for theoretical MD calculated data by Harafuji et al.[9], ★ denotes melting temperatures obtained in this report, —— is for the equilibrium curve from Figure 3, —— $T_m(P)$ curve obtained by DR[11] model

based on $T_m(P)$ data by Harafuji et al.[9] and from this paper, —— hypothetical equilibrium line between GaN and its constituents, ▲ ★ – GaN stable crystals, △ ☆ – GaN decomposed to $(Ga+N)_{liq.}$ and $N_2$.

The obtained $T_m(P)$ curve shows a maximum at ca. 22 GPa, associated with the cross over from $dT_m/dP > 0$ to $dT_m/dP < 0$ domain. It is worth recalling that for GaN Van Vechten[7] assumed the change of the coordination number (CN) at melting from CN=4 for Wurtzite phase to CN ≈ 6 for the liquid state and linked it to $dT_m/dP < 0$. On the other hand Harafuji et al.[9] obtained for GaN that $dT_m/dP > 0$ and showed that CN at melting does not change substantially and remains close to 4 in the liquid state. Hence one can expect that the predicted maximum of $T_m(P)$ maybe associated with the gradual change of the coordination number from[9] CN≈4 (Low Density Liquid – LDL) to[9] CN≈6 (High Density Liquid – HDL) in the liquid state on increasing pressure. It is support by the recent finding for liquid germanium at normal pressure where in spite of the dominating coordination number CN≈6 local fluctuations related to CN≈4 were observed[8]. We correlate this behaviour with the fact that for germanium a maximum of $T_m(P)$ dependence is predicted in the negative pressure, already at $P \sim -0.3 GPa$.[26]

It is notable that within Van Vechten[7] theory the similarity of phase diagrams or GaN and c-BN was stressed. For both semiconductors $dT_m/dP < 0$ was predicted. However, two decades ago, for c-BN the experimental evidence showing $dT_m/dP > 0$ and the maximum of $T_m(P)$ at ca. 3 GPa was reported.[28,29] This result resembles the behaviour found for GaN in the given report.

Figure 5b presents also the experimental equilibrium curve (EQ) for GaN and its constituents $GaN(solid) \rightarrow Ga + N(liquid\_solution) + N_2(gas)$. The crossing of its linear extrapolation (dashed blue line) and the $T_m(P)$ curve occurs at $P \approx 12$ GPa and $T \approx 4000$ K.

For lower pressures the decomposition of GaN is expected prior to melting. The cross-section of melting curve $T_m(P)$ and EQ curve can be seen as "quasi triple point" at which solid and liquid GaN coexist with the mixture of gaseous $N_2$ and liquid Ga+N solution.

Rresults presented in Figure 5b are in disagreement with the evidence by Utsumi et al.[10], who claimed that the decomposition of GaN can be observed only for pressures below $P \approx$ 5.8 GPa whereas for higher pressures solely melting with $dT_m/dP \approx 0$ takes place. We would like to stress that in our *HP-HT* experiments samples were ca. 100x larger volume than ones used by Utsumi et al.[10], what essentially reduced the impact of parasitic artefacts.

## 3. Conclusions

*HP-HT* studies of GaN made it possible to increase the solubility of nitrogen in gallium up to 17% . Notwithstanding this is still the congruent melting at which concentration of N atoms in liquid reaches 50%. Up to $P = 9$ GPa and $T = 3400$ K solely the decomposition was observed. The scaling analysis of nitrogen solubility in gallium allowed for determining GaN melting curve $T_m(P)$ in the domain where it is hidden by earlier decomposition. The obtained $T_m(P)$ experimental data can be well portrayed by DR equation with pressure invariant parameters, what validates the extrapolation beyond the experimental range of pressurea. Worth indicating is the maximum of $T_m(P)$ at ca. 22 GPa what may be associated with an increase of the coordination number in liquid GaN from $CN \approx 4 \to CN \approx 6$.

Notable is the very good agreement with results obtained by Harafuji et al.[9] MD simulations as well as disagreement both with theoretical predictions of Van Vechten[7] and the experimental evidence by Utsumi et al.[10] . All these results are shown in Fig. 6.

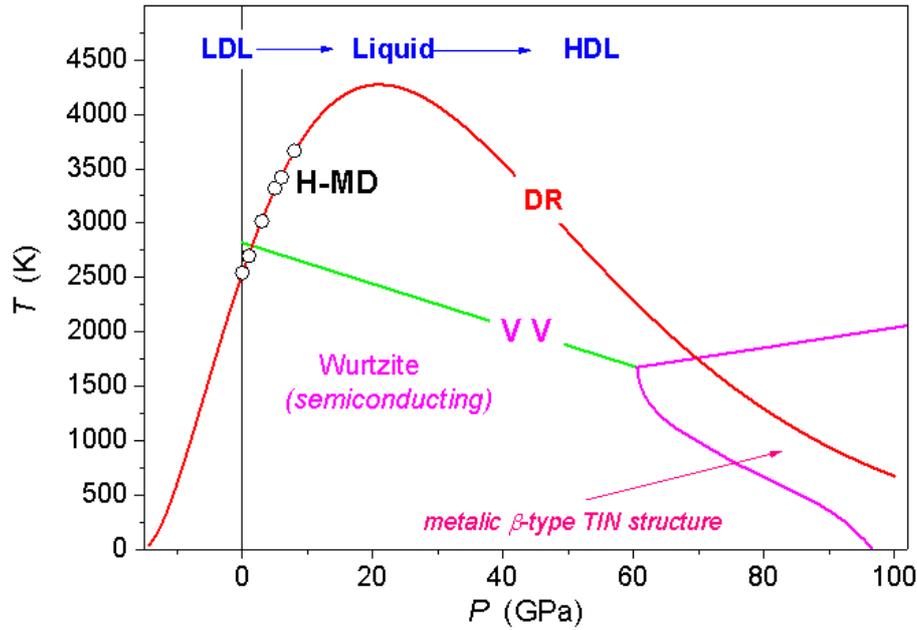

**Figure 6** Model melting curves for GaN. $T_m(P)$ curves and data for GaN predicted: —— by Van Vechten[7] theory (VV), ○ by Harafuji et al.[9] MD simulations (H-MD) and —— the DR[11] relation (eq.(3)). The latter can also predict the extension of $T_m(P)$ curve into negative pressures domain.

Concluding, this paper proposes the new way of analysis for determining $T_m(P)$ evolution in the *P-T* domain for GaN and similar materials which decompose prior to melting as well as solves the long standing puzzle emerging due to contradictory predictions of refs.[7,9,10]. The excellent agreement of obtained results with Harafuji et al.[9] MD simulation reinforces this fundamental approach as the tool for reaching the insight into GaN properties. Results obtained also open a new avenue for modelling relevant technological processes in GaN, particularly related to the crystal growth.

## 4. Experimental

The high pressure – high temperature experiments had been performed using toroidal high pressure (*HP*) apparatus[14] and modified sample assembly for the industrial growth of diamonds. The anvils of the apparatus combine the toroidal cave around central one in form of spherical segments 30 mm in diameter. Sample assembly for experiments at very high temperatures up to 3500 K (*P*=6–9 GPa) consist of graphite capsule surrounded by Ta capsule which were placed into CsCl sleeve and then into cylindrical graphite heater (see Fig. 1). Estimation of the sample temperature based on heat calibration of the *HP* cell (sample temperature vs heater power) which was found using known melting temperature points weakly dependent on pressure. These standard $T_m$ points were determined to be 2473 K for contact eutectic melting reaction in Mo-C system i.e. point of α-β$Mo_2C$ eutectic[30], and 2947 K (at pressure 8 GPa) for pure molybdenum melting by Fateeva et al.[31]. The calibration of pressure at high temperature was based on observation of diamond graphitization in presence of Co melt as an activating agent of the transformation. Since overheating of the diamond in this case is not possible the sample pressure is close to equilibrium value conformed to the given temperature. For analysis extrapolated dependence of diamond-graphite equilibrium line by Kennedy and Kennedy[32] $P(GPa) = 1.26 + 0.0025 \cdot T(K)$ was used. For example, at compression force of $F = 1750$ tons diamond graphitization in sample volume was observed starting at temperature 2573 K which correspond to pressure equal to 7.7 GPa. We estimated the accuracy of *P, T* parameters determination to be about ±5 % for pressure and ±100 K for temperature. In our experiments the GaN crystals (platelets) grown by HVPE method[15] were surrounded by GaN powder obtained by milling of such crystals to grain dimensions below 80 *μm* (Fig. 1b). The prepared samples (platelets and the powder) were located in the cylindrical graphite capsule with 1 mm thick walls, which was subsequently placed in the tantalum capsule 0.2 mm thick and with Ø 13 mm (diameter), $h = 7$ mm (height). It is shown in Figure 1b. Typically, three GaN platelets about 0.4 mm thick were placed inside the capsule. The high pressure toroid-type apparatus DO-043 supported by 2 000 tones hydraulic press was used[14]. Experiments consisted of three stages: (i) pressure was increased up to the selected value during 1 min, (ii) the second stage consisted of three parts, namely (a) heating power was increased to the value corresponding to the planned temperature during $t = 2$ s and the stable value of temperature was reached in 10 s, (b) power was kept constant during 40 s, (c) next the power was decreased during 10 s and finally, (iii) pressure was released during 1 min.

After each *HP-HT* run the capsule with the sample was removed from the pressure cell and properties of the sample were analyzed by HR XRD and powder XRD methods, SEM, optical microscopy and compared with properties of starting materials.

**Table I.** Analysis of solubility data.

|   | $P$ (GPa) | $T$ (K) | $S$, (atomic fraction) | $T/T_m$ | $T_m$ (K) | $T_m(P)$ via eq. (3) |
|---|---|---|---|---|---|---|
|   | 1 | 2 | 3 | 4 | 5 | 6 |
| 1 | 0.81 | 1720 | 0.001 (ref.[16]) | 0.62 | 2775 | 2675 |
| 2 | 1.50 | 1850 | 0.005 (ref.[16]) | 0.65 | 2845 | 2790 |
| 3 | 6 | 2800 | 0.037 | 0.77 | 3635 | 3420 |
| 4 | 8 | 3150 | 0.110 | 0.87 | 3620 | 3655 |
| 5 | 9 | 3400 | 0.170 | 0.92 | 3700 | 3755 |
| 6 | 20 | 3200 | 0.050 (ref.[18]) | 0.79 | 4050 | 4270 |

Samples related to type I results (GaN stable) after *HP-HT* process were characterized by HR XRD. This characterization has been made to check if melting to GaN during type I experiments occurred. The symmetrical reflection X-ray rocking curves (omega scan) of reference crystals and crystals after *HP-HT* processes were measured using high resolution X-ray diffractometer Philips X'Pert Pro, equipped with a four reflection Bartels monochromator. The X-ray beam had width of 0.1 mm in diffraction plane. The rocking curves for reflection (3 0 0) were measured on the cross-section of GaN crystals before and after *HP-HT* treatment.

Samples of the newly grown GaN platelets (type II decomposition results) were characterized by powder XRD and compared with standard powder XRD for GaN. The position of XRD peaks for standard diffraction pattern and newly grown platelets are identical, confirming that newly grown crystals are gallium nitride. The SEM picture showed that newly grown crystals have form of hexagonal platelets typical for GaN grown from solution[13].

**Acknowledgements**

The research was supported by FP7 START Project "*Boosting EU-Ukraine cooperation in the field of Superhard Materials*" ref. 295003, Polish grant MNiSW "*Kwantowe nanostruktury półprzewodnikowe do zastosowań w biologii i medycynie*" POIG 01.01.02-008/08-00 and NCN grant (Poland) ref. UMO-2012/05/N/ST3/02545. SJR was supported by National Centre for Science (Poland) grant 2011/03/B/ST3/02352. The Authors would like to thank Jörg Neugebauer and Martin Albrecht for the critical discussion, Adam Presz and Igor Dzięcielewski for SEM tests and Jan Weyher for etching samples.

**FIGURES**

**Figure 1**   Toroid type apparatus for HP-HT studies: (a) photo of the pressurized part of the apparatus with the cell, and (b) the scheme of cross-section of *HP-HT* cell with graphite heater and the sample capsule.

**Figure 2**   Optical microscopy and SEM views of GaN samples after *HP-HT* treatments revealing two types of behaviour. Type I: (a) there are no visible changes in the shape of GaN crystals and (b) there are hallmarks of GaN powder sintering. Type II: (c) the whole sample is decomposed and gallium droplets are visible, and (d) after removal (etching) of gallium newly grown crystals appear.

**Figure 3**   GaN decomposition (EQ) curve. Experimental equilibrium curve for the solid GaN, liquid Ga+$N_{atomic}$ solution and gaseous $N_2$ in the pressure–temperature plane. ▲ ★ – GaN stable crystals, △ ☆ – GaN decomposed to $(Ga+N)_{liq.}$ and $N_2$, ▬ the "guide for eyes" interpolation equilibrium curve.

**Figure 4**   Liquidus equilibrium (EQ) curves for GaSb, GaP and GaN for different pressures (a) and scaled liquidus curves for these compounds (b). ▬ is for GaSb, ▬ for GaP (based on data from ref.[21]) and ▬ for GaN from this work. ○ – Experimental solubility values obtained by Grzegory et al.[11], ○ – theoretical solubility value obtained by Nord et al.[18], ○ – experimental solubility values from this work.

**Figure 5** Pressure dependence of GaN melting and GaN *P-T* equilibrium curve (EQ). a, shows results of the derivative based analysis, via eqs. (4) showing the failure of the SG[22] eq. (3), which validity ought to appear via the linear domain. Such behaviour is presented in the inset, where DR[11] eq. (3) indicated the possibility of parameterization via DR[11] eq. (4) is shown. The linear regression fit yields optimal values of all relevant coefficient ($\pi, b, c$) for eq. (3). $T_m(P)$ curve developed on this base is shown in the main part of b. ○ is for theoretical MD calculated data by Harafuji et al.[9], ★ denotes melting temperatures obtained in this report, ▬ is for the equilibrium curve from Figure 3, ▬ $T_m(P)$ curve obtained by DR[11] model based on $T_m(P)$ data by Harafuji et al.[9] and from this paper, ▬ hypothetical equilibrium line between GaN and its constituents, ▲ ★ – GaN stable crystals, △ ☆ – GaN decomposed to $(Ga+N)_{liq.}$ and $N_2$.

**Figure 6** Model melting curves for GaN. $T_m(P)$ curves and data for GaN predicted: ▬ by Van Vechten[7] theory (VV), ○ by Harafuji et al.[9] MD simulations (H-MD) and ▬ the DR[11] relation (eq.(3)). The latter can also predict the extension of $T_m(P)$ curve into negative pressures domain.

## Authors contributions

S. P., B. S., I. G., I. P. and V. T. – design of *HP-HT* experiments. S. P., B. S., I. P. and D. S. – preparation and conducting the *HP-HT* experiments, S. P., B. S. and I. P. – preparation of samples and their characterisations before and after *HP-HT* experiments. S. P., B. S., S. J. R., I. G., I. P. and V. T. – analysis and discussion of the experimental results. S. P., B. S., S. J. R. and I. G. – writing the manuscript.